\documentclass[journal]{IEEEtran}
\usepackage{color}
\usepackage{enumitem}
\usepackage{array}
\usepackage{subfigure}

\usepackage{cite}

\ifCLASSINFOpdf
  \usepackage[pdftex]{graphicx}
  \graphicspath{{./fig/}}
\else
\fi
\usepackage{amsmath,amssymb}

\usepackage{stfloats}
\begin{document}
%
\title{Enhanced Atmospheric Turbulence Resiliency with Successive Interference Cancellation DSP in Mode Division Multiplexing Free-Space Optical Links}
%
%
%

\author{Yiming~Li,
        Zhaozhong~Chen,
        Zhouyi~Hu,~\IEEEmembership{Member,~IEEE,}
        David~M.~Benton,
        Abdallah~A.~I.~Ali,
        Mohammed~Patel,
        Martin~P.~J.~Lavery,
        and~Andrew~D.~Ellis,~\IEEEmembership{Fellow,~Optica}

\thanks{Manuscript received June 00, 2022; revised June 00, 2022. Research supported by EPSRC under grant number EP/T009047/1, EP/T009012/1, EP/S003436/1, EP/S016171/1, and the European Union’s Horizon 2020 research and innovation programme under the Marie Skłodowska-Curie grant agreement No. 713694.}
\thanks{Y.~Li, Z.~Hu, D.~Benton, A.~Ali, M.~Patel, and A.~Ellis are with Aston Institute of Photonic Technology, Aston University, Birmingham, B4 7ET, UK (e-mail: y.li70@aston.ac.uk; z.hu6@aston.ac.uk; d.benton@aston.ac.uk; a.ali81@aston.ac.uk; m.patel70@aston.ac.uk; andrew.ellis@aston.ac.uk).}
\thanks{Z.~Chen, and M.~Lavery are with James Watt school of engineering, University of Glasgow, Glasgow, G12 8QQ, UK (e-mail: Zhaozhong.Chen@glasgow.ac.uk; Martin.Lavery@glasgow.ac.uk).}
}

%
%

\markboth{Journal of Lightwave Technology,~Vol.~00, No.~0, August~2022}%
{Shell \MakeLowercase{\textit{et al.}}: Bare Demo of IEEEtran.cls for IEEE Journals}
%



\maketitle

\begin{abstract}
We experimentally demonstrate the enhanced atmospheric turbulence resiliency in a 137.8\,Gbit/s/mode mode-division multiplexing free-space optical communication link through the application of a successive interference cancellation digital signal processing algorithm. The turbulence resiliency is further enhanced through redundant receive channels in the mode-division multiplexing link.
The proof of concept demonstration is performed using commercially available mode-selective photonic lanterns, a commercial transponder, and a spatial light modulator based turbulence emulator. In this link, 5 spatial modes with each mode carrying 34.46\,GBaud dual-polarization quadrature phase shift keying signals are successfully transmitted with an average bit error rate lower than the hard-decision forward error correction limit. As a result, we achieved a record-high mode- and polarization-division multiplexing channel number of 10, a record-high line rate of 689.23\,Gbit/s, and a record-high net spectral efficiency of 13.9\,bit/s/Hz in emulated turbulent links in a mode-division multiplexing free-space optical system.
\end{abstract}


%
\IEEEpeerreviewmaketitle

\section{Introduction}
\label{sec:i}
%
%
%
%

\IEEEPARstart{F}{ree} space optical (FSO) communication is a promising wireless communication technique for a wide range of applications. FSO systems can provide a very high unlicensed bandwidth for data transmission at more than one hundred Gigabits per second \cite{ren2016experimental},
and can also provide extremely long transmission distance, inherent security, and robustness to electromagnetic interference \cite{lange2006142,khalighi2014survey}.

Mode-division multiplexing (MDM) can be used in FSO communication systems to increase the throughput or mitigate the impact of atmospheric turbulence \cite{wang2012terabit,fontaine2019digital}. A popular basis for MDM is orbital angular momentum (OAM) modes, which can be decomposed into the rotational symmetric complete orthogonal Laguerre–Gaussian (LG) modal basis set \cite{willner2021perspectives,qu2016500,ren2014adaptive}. To further increase the transmission capacity, it is preferable to employ a complete orthogonal mode set, such as LG modes \cite{trichili2019communicating}, where orthogonality guarantees negligible crosstalk, and the rotational symmetry guarantees good compatibility with conventional optical lenses and fiber based delivery with relatively low modal crosstalk \cite{bruning2015overlap}.

Atmospheric turbulence can significantly reduce the system performance by introducing scintillation, beam wandering, beam spreading, angle-of-arrival fluctuations, and phase fluctuations \cite{andrews2005laser}. The performance degradation is even more pronounced in a MDM FSO system as it can introduce the inter-mode crosstalk \cite{qu2016500,li2019mitigation}.

Adaptive optics (AO) may be used to mitigate atmospheric turbulence effects 
\cite{ren2014adaptive,chen2016demonstration,weyrauch2001microscale},
and may be assisted by a coaxial Gaussian beam as reference for MDM FSO systems \cite{ren2014adaptive}. However, the AO systems can not fully compensate for the turbulence effects and residual inter-mode crosstalk remains \cite{ren2014adaptive,chen2016demonstration}, requiring a tradeoff between control bandwidth and compensation errors \cite{weyrauch2001microscale}.

Multiple-input multiple-output (MIMO) digital signal processing (DSP) is another approach to mitigate the negative effects of atmospheric turbulence in the MDM FSO systems \cite{huang2014crosstalk,ren2016atmospheric}. MIMO DSP shifts the complexity from the optical to the digital domain, enabling rapid feed-forward channel state information (CSI) extraction. Most reported MIMO algorithms in optical communications are based on a minimum mean square error (MMSE) MIMO equalizer \cite{huang2014crosstalk,ren2016atmospheric,van2014time,rademacher2021peta,shibahara2021mimo}. This works well in few-mode fibre (FMF) systems where the channel matrix is almost unitary \cite{van2014time,shibahara2021mimo,rademacher2021peta}, but introduces significant performance degradation in fading channels \cite{proakis2001digital,huang2014crosstalk,ren2016atmospheric}.

It is widely accepted in the radio frequency wireless MIMO communication systems that adding redundant receive channels can provide extra diversity, which is beneficial to the bit error rate (BER) performance \cite{proakis2001digital}. To date, in MDM FSO systems, diversity gain has only been considered in special cases such as single-input multiple-output (SIMO) systems or systems with multiple spatially separated apertures \cite{fontaine2019digital,ren2016atmospheric}. The generalized MIMO algorithms with redundant receive channels have not been applied to the MDM FSO systems to combat turbulence induced fading yet.

In this paper, we experimentally demonstrated enhanced atmospheric turbulence resiliency in MDM FSO communication systems. In particular: (1) We validate for the first time that successive interference cancellation (SIC) MIMO decoding provides enhanced turbulence resiliency;
(2) We provide the first demonstration that a generalized MIMO algorithm with redundant receive channels provides enhanced turbulence resiliency by comparing different MIMO systems (e.g. \(10 \times 10\), \(10 \times 12\), and \(6 \times 6\) to \(6 \times 12\));
(3) We employ commercial components, including mode selective photonic lanterns (MSPLs) and coherent transponders to demonstrate practical applicability.
As a result, we have demonstrated a record high independent channel number of 10, line rate of 689.23\,Gbit/s, and net spectral efficiency of 13.9\,bit/s/Hz in emulated turbulent MDM FSO links.

The remainder of this paper is organized as follows: the experimental setup is given in Section~\ref{sec:es}, Section~\ref{sec:tg} proposes the selection of phase screens for turbulence generation, Section~\ref{sec:mda} proposes the SIC MIMO DSP with redundant receive channels for enhanced turbulence resiliency, Section~\ref{sec:er} provides the experimental results, and Section~\ref{sec:c} summarizes the key advantages of the proposed MDM FSO system.

\section{Experimental Setup}
\label{sec:es}

\begin{figure*}[htb]
  \centering
  \includegraphics[width=7.1in]{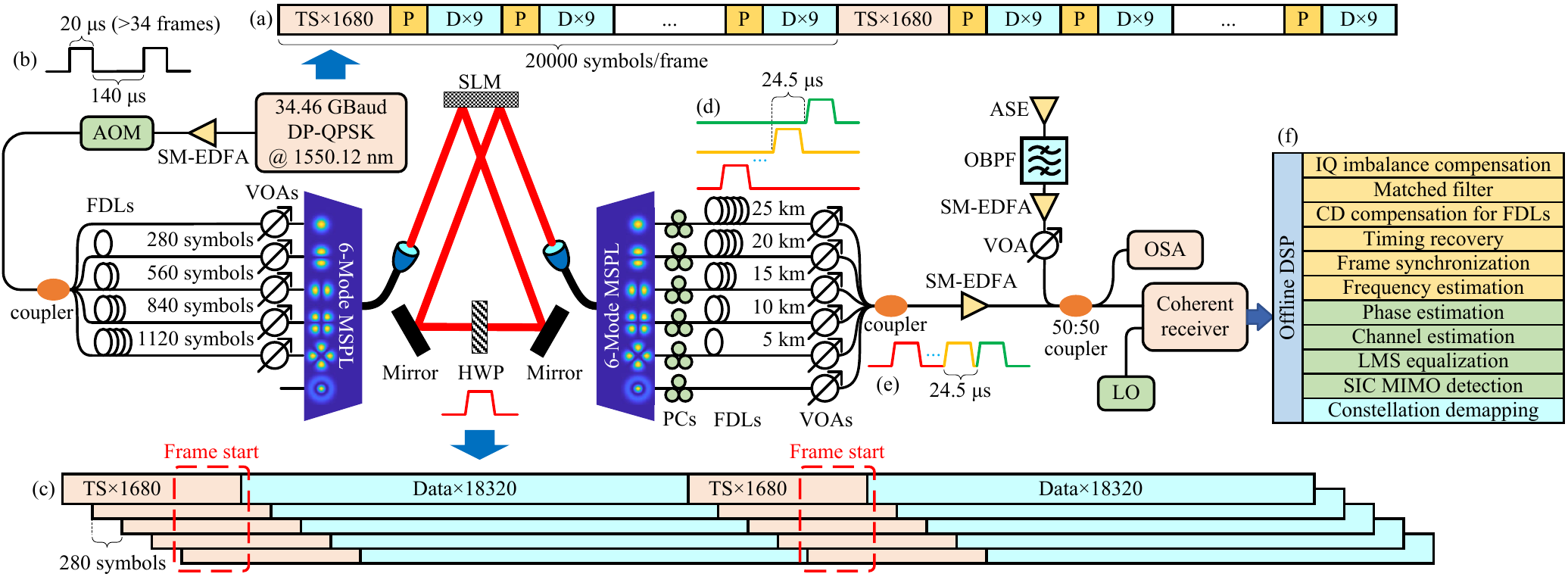}
  \caption{The experimental setup. DP-QPSK: dual-polarization quadrature phase shift keying; TS: training sequence; P: pilot; D: data; SM-EDFA: single-mode erbium-doped fiber amplifier; AOM: acousto-optic modulator; FDLs: fibre delay lines; VOAs: variable optical attenuators; MSPL: mode-selective photonic lantern; SLM: spatial light modulator; HWP: half-wave plate; PCs: polarization controllers; ASE: amplified spontaneous emission; OBPF: optical bandpass filter; OSA: optical spectrum analyzer; DSP: digital signal processing;
  CD: chromatic dispersion; LMS: least mean squares; SIC: successive interference cancellation; MIMO: multiple-input multiple-output. (a) Frame structure generated by Ciena transponder; (b) signal burst after AOM; (c) delayed frame structure after transmitter FDLs; (d) time-division multiplexing (TDM) signal after receiver FDLs; (e) TDM signal after receiver coupler; (f) offline DSP.}
  \label{fig:system}
  \end{figure*}

The experimental setup is shown in Fig.~\ref{fig:system}.
At the transmitter, a Ciena WaveLogic 3 transponder with a 39.385\,GSa/s onboard arbitrary waveform generator (AWG) was used to generate the 34.46\,GBaud dual-polarization quadrature phase shift keying (DP-QPSK) signals at 1550.12\,nm. As shown in Fig.~\ref{fig:system}(a), the signals had a pilot-aided frame structure with a frame length of 20,000 symbols. Each frame had a training sequence of \(1680\) symbols, and a pilot symbol for every 9 data symbols in the payload. The training sequence and the pilots were generated by a random permutation of balanced QPSK symbols.
The data symbols were generated from a PRBS-15 pseudo random binary sequence (PRBS). The generated signals were shaped by a root-raised cosine (RRC) filter with a roll-off factor of 0.1. The signals were then amplified by a single-mode erbium-doped fibre amplifier (SM-EDFA). In order to emulate the use of independent transmitters and receivers, at the transmitter, the resultant optical signal was passed through an acousto-optical modulator (AOM), which was operated at a period of 160\,\textmu s and a duty cycle of 12.5\%, generating a 20\,\textmu s signal burst (Fig.~\ref{fig:system}(b), more than 34 frames).
The signals were then split into 5 copies by a fibre coupler and then each copy was delayed by a variable fibre delay line (FDL) with lengths of 0, 280, 560, 840, and 1120 symbols, respectively to generate decorrelated signals for different modes (Fig.~\ref{fig:system}(c)), and the decorrelation length between adjacent modes is 1/6th of the training sequence length. The decorrelated signals were then connected to 5 variable optical attenuators (VOAs) to compensate for the mode-dependent loss at the transmitter side to guarantee that each transmitted mode contributed approximately the same power at the receiver side. The attenuated signals were finally connected to the five lowest order linearly polarized (LP) modes of a 6-mode MSPL for mode multiplexing and then coupled into free space using a collimator with a focal length of 18.4\,mm and a lens diameter of $D=8.4$\,mm.

In the FSO channel, we employed a spatial light modulator (SLM) with \(1920 \times 1152\) pixels to emulate indefinite number of independent turbulence realizations. The size of each pixel is $9.2 \times 9.2$\,\textmu $\text{m}^2$. 
In order to modulate the both polarization components of the signal, we passed the collimated light through the SLM twice (red paths in Fig.~\ref{fig:system}). Simulated phase screen were displayed on the left and right half of the SLM. The distance between the collimators and the SLM, the mirrors and the SLM, and the two mirrors were \textasciitilde 50\,cm, \textasciitilde 50\,cm, and \textasciitilde 20\,cm, respectively. During the first pass, the horizontal polarization component was modulated by turbulence phase screen.
A half-wave plate (HWP) located in-between the first and second pass swapped the horizontal and vertical components. 
And then the unaffected component in the first pass was modulated by the same turbulence phase screen during the second pass.
The selection of phase screens for turbulence generation will be detailed in Section~\ref{sec:tg}.

At the receiver side, the free-space beam was coupled into a FMF by a second identical $D=8.4$\,mm diameter FSO coupler with a focal length of 18.4\,mm. 
The received signals were then demultiplexed by a second 6-mode MSPL at the receiver side. 6 polarization controllers were connected to the output of the MSPL to balance the received power in different polarization states. Afterwards, the 6 signals were delayed by the 25\,km, 20\,km, 15\,km, 10\,km, 5\,km, and 0\,km FDLs to enable the time-division multiplexing (TDM) receiver \cite{li2022demonstration,hu2022single}. As shown in Fig.~\ref{fig:system}(d), the FDLs generated an approximately 24.5\,\textmu s delay between adjacent modes, which was slightly longer than the signal burst. 6 VOAs were then connected to the output of the FDLs to compensate for the mode-dependent loss of the devices at the receiver side. The TDM signals were then coupled into one single-mode fibre (Fig.~\ref{fig:system}(e)) and amplified by another SM-EDFA.
Afterwards, the variable optical noise was loaded by a combination of an amplified spontaneous emission source, an optical bandpass filter, a SM-EDFA, a VOA, and a 50:50 coupler. The average optical signal-to-noise ratio (OSNR) was then measured by an optical spectrum analyzer to enable OSNR scanning.
The resultant signal was received by a standard coherent receiver with a 50\,GSa/s sampling rate and 23\,GHz bandwidth oscilloscope. Finally, the received digital traces were demodulated and decoded by the offline DSP (Fig.~\ref{fig:system}(f)).

\section{Turbulence Generation}
\label{sec:tg}

In this work, we emulated the beam propagation in a thin phase turbulent channel representing a channel approximately 400-500\,m in length, using a single aberration inducing phase screen implemented with a SLM to emulate each instantiation of the turbulent channel. The screens used were generated by implementing a power spectrum inversion method, in which phase screens are randomly generated by performing a Fast Fourier Transform (FFT) of associated spatial frequency power spectrum. The power spectrum that is commonly used for atmospheric turbulence is the von Kármán spectrum \cite{andrews2005laser}, that has Fourier series coefficients, $c_{n,m}$, in the form,
\begin{equation}
\begin{aligned}
  {c_{n,m}} = & w(n,m)\frac{{\sqrt {0.023} }}{{{L}}}{\left( {\frac{2}{{{r_0}}}} \right)^{^{\frac{5}{6}}}}{\left( {f_{{x_n}}^2 + f_{{y_m}}^2 + \frac{1}{{L_0^2}}} \right)^{ - \frac{{11}}{{12}}}} \\ 
  & \times {\left\{ {\exp \left[ { - \left( {f_{{x_n}}^2 + f_{{y_m}}^2} \right){{\left( {\frac{{2\pi {l_0}}}{{5.92}}} \right)}^2}} \right]} \right\}^{\frac{1}{2}}} ,\\ 
\end{aligned}
\end{equation}
where $w(n, m)$ is the random array that obey complex circular Gaussian statistics with zero-mean and unit-variance. 
$L=8.832$\,mm is the length of the $960 \times 960$ pattern. $r_0$ is the Fried's parameter.
The spatial frequency grid is determined by parameters $f_{x_n}=n/L$ and $f_{y_m} =m/L$. Further, this power spectrum is influenced by the physical channel parameters that determine the largest, $L_0$ to smallest, $l_0$ eddies sizes present in a turbulent channel.

Turbulence-induced phase $\phi(x, y)$ can be generated by considering discrete x- and y-directed spatial frequencies denoted as $f_{x_n}$ and $f_{y_m}$, respectively. To accurately represent modes like tilt aberration, the sub-harmonic can be employed
\cite{Lane1992SimulationScreen,cheng2021}. A \(3 \times 3\) 
grid of 2D arrays of spatial frequencies is used, where the frequency grid spacing for each value of $p$ is \({1 \mathord{\left/
 {\vphantom {1 {\left( {3^p L} \right)}}} \right.
 \kern-\nulldelimiterspace} {\left( {3^p L} \right)}}\), and the corresponding Fourier series coefficient is $c_{n,m,p}$. A specific phase screen, $\phi(x, y)$ can then be computed as a sum of $N_p + 1$ independent screens using the formula
\begin{equation}
\begin{aligned}
  \phi (x,y) = & \sum\limits_{n = - \infty }^{ + \infty } {\sum\limits_{m = - \infty }^{ + \infty } {{c_{n,m}}\exp \left[ {j2\pi \left( {{f_{{x_n}}}x + {f_{{y_m}}}y} \right)} \right]} } \hfill \\
   + & \sum\limits_{p = 1}^{{N_p}} {\sum\limits_{n = - 1}^{ 1} {\sum\limits_{m = - 1}^{ 1} {{c_{n,m,p}}\exp \left[ {j2\pi \left( {{f_{{x_n}}}x + {f_{{y_m}}}y} \right)} \right]} } } \hfill .\\ 
\end{aligned}
\end{equation}

\begin{figure}[tb]
    \centering
    \subfigure{
      \centering
      \includegraphics[width=1.2in]{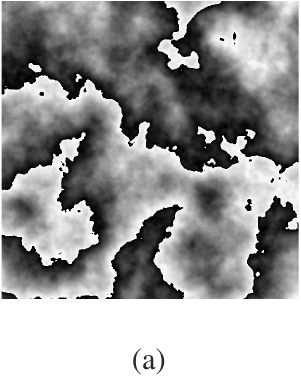}
    }
    \subfigure{
      \centering
      \includegraphics[width=2in]{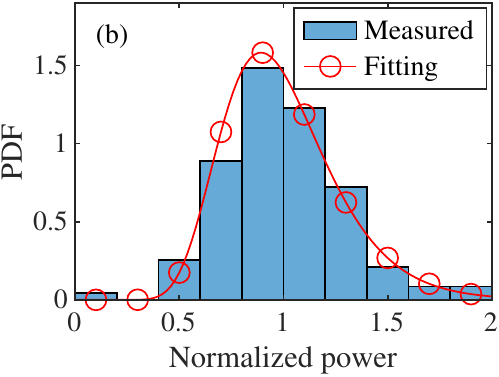}
    }
    \subfigure{
      \centering
      \includegraphics[width=1.2in]{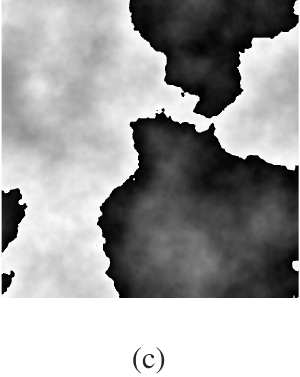}
    }
    \subfigure{
      \centering
      \includegraphics[width=2in]{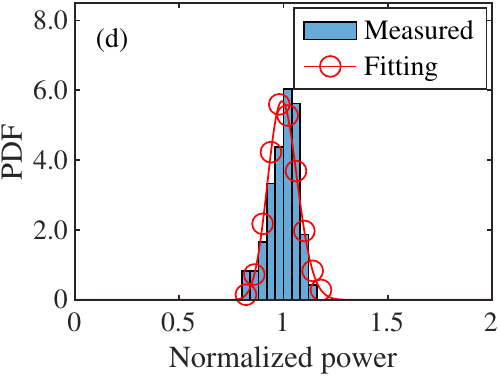}
    }
    \caption{The typical turbulence realizations (\(L_0=10\,\text{m}\), \(l_0=0.1\,\text{mm}\)). PDF: probability density function. (a) The typical pattern of the stronger turbulence (\(r_0=0.8\,\text{mm}\)); (b) The power distribution of the stronger turbulence (\(\sigma_I^2=0.079\)); (c) The typical pattern of the weaker turbulence (\(r_0=3.0\,\text{mm}\)); (d) The power distribution of the weaker turbulence (\(\sigma_I^2=0.0050\)).}
    \label{fig:typical_pattern}
  \end{figure}

Each computed screen provides an individual instance of turbulence and can vary in strength, where the full impact of a particular turbulence strength can only be realized by considering an appropriate number of instances over time. It is generally considered that turbulence is approximately frozen at a time base close to the coherence time, commonly called the Greenwood frequency \cite{cheng2021}, of the atmosphere and can vary between 60\,Hz to 1\,kHz depending on the particular environment condition. Therefore, if a channel measurement is performed suitably faster than this Greenwood frequency, which is the typical case for the channel estimation algorithm in our system, then one single instance of turbulence needs to be considered when the channel mitigation strategies are applied.
We considered two sets of 120 independent patterns, both with \(L_0=10\,\text{m}\) and \(l_0=0.1\,\text{mm}\), for stronger (\(r_0=0.8\,\text{mm}\), \(D/r_0=10.5\)) and weaker (\(r_0=3.0\,\text{mm}\), \(D/r_0=2.8\)) turbulence, respectively. Example patterns are shown in Fig.~\ref{fig:typical_pattern}(a) and Fig.~\ref{fig:typical_pattern}(c), whilst Fig.~\ref{fig:typical_pattern}(b) and Fig.~\ref{fig:typical_pattern}(d) show the normalized power probability density function (PDF) at the receiver (before MSPL), which agree well with normalized lognormal curve fits corresponding to scintillation indexes (\(\sigma_I^2\)) of 0.079 and 0.0050, respectively \cite{andrews2005laser}.


\section{MIMO DSP Algorithm}
\label{sec:mda}
\subsection{Preliminaries}
\label{sec:p}

\begin{figure}[tb]
  \centering
  \includegraphics{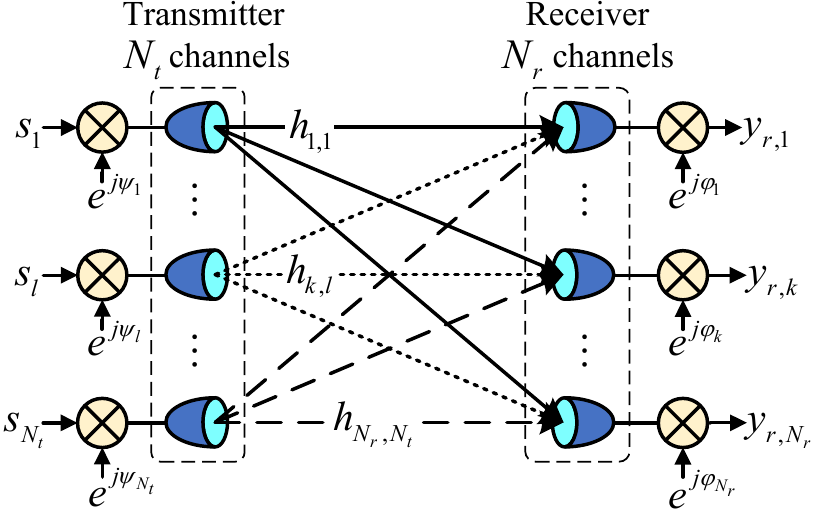}
  \caption{Schematic diagram for TDM MDM optical communication systems.}
  \label{fig:Schematic_Diagram}
  \end{figure}

As shown in Fig.~\ref{fig:Schematic_Diagram},
a generalized mathematical model for the TDM MDM optical communication system with \(N_t\) transmit channels and \(N_r\) receive channels can be given as
  \begin{equation}\label{equ:p.1}
  {{\mathbf{y}}_r} = {\mathbf{\Phi H\Psi s}} + {{\mathbf{n}}_r},
  \end{equation}
where \({\mathbf{y}}_r\) is the \(N_r \times 1\) received signal vector, \(\mathbf{H}\) is the \(N_r \times N_t\) channel matrix, \({\mathbf{s}}\) is the \(N_t \times 1\) transmitted signal vector, \({\mathbf{n}}_r\) is the \(N_r \times 1\) independent and identically distributed (i.i.d.) circularly-symmetric complex additive white Gaussian noise (AWGN) vector, where the expectation and the variance of each element are 0 and \(N_0\), respectively.
\({\mathbf{\Psi }} = {\text{diag}}\left( {{e^{j{\psi _1}}}, \cdots ,{e^{j{\psi _{N_t}}}}} \right)\) and \({{\mathbf{\Phi }}} = {\text{diag}}\left( {{e^{j{\varphi _{1}}}}, \cdots ,{e^{j{\varphi _{{N_r}}}}}} \right)\) are the diagonal phase matrices of the transmit and receive channels, where \(\psi _l\) and \(\varphi _k\) are the phase noise 
at the \(l^{th}\) transmit channel and 
the \(k^{th}\) receive channel, respectively.
If the diagonal elements of \(\mathbf{\Phi }\) and \(\mathbf{\Psi}\) are approximately identical, corresponding to a single transmit laser and a local oscillator, both with phase matched fiber paths, then MIMO equalization may be carried out prior to phase estimation \cite{van2014time}. However, the approximately identical phase should be conserved by carefully splitting and delaying the LO beam or using multiple oscilloscopes at the receiver \cite{rademacher2021peta}. This extra complexity can be reduced by employing carrier-asynchronous algorithms to combat decorrelated LOs, but prior attempts have suffered from iterative phase and channel estimation structure \cite{shibahara2021mimo}. Moreover, inherent difficulties exist in the above mentioned DSPs when employing advanced MIMO decoding algorithms such as SIC.

To accommodate these practical realities, we propose a new carrier-asynchronous DSP structure, which is shown in Fig.~\ref{fig:system}(f) (green text), to separate the MIMO decoder from the phase estimation, the channel estimation, and the ISI equalization. By doing so, redundant receive channels and advanced MIMO decoding algorithms such as SIC can be supported.

The phase and channel estimation algorithm is fully detailed in one of our previous work \cite{li2021crlb}. Noting the fact that the phases are well conserved at the transmitter side, we can set \(\mathbf{\Psi}\) to a \(N_t \times N_t\) identity matrix as reference. Therefore, \eqref{equ:p.1} can be simplified to
  \begin{equation}\label{equ:p.2}
  {\mathbf{y}}_r = {\mathbf{\Phi H s}} + {\mathbf{n}}_r.
  \end{equation}
After the phase and channel estimation, we can obtain \(\mathbf{\hat \Phi}\) and \({\mathbf{\hat H}}\), which are the estimations of \(\mathbf{\Phi}\) and \(\mathbf{H}\), respectively. And the phase information can be cancelled out as
  \begin{equation}\label{equ:p.3}
  {\mathbf{y}} = {{\mathbf{\hat \Phi }}^H}{{\mathbf{y}}_r} \approx {\mathbf{Hs}} + {{\mathbf{\hat \Phi }}^H}{{\mathbf{n}}_r} = {\mathbf{Hs}} + {\mathbf{n}},
  \end{equation}
where \({\left( \cdot \right)^H}\) is the Hermitian adjoint operator and \({\mathbf{n}} = {{\mathbf{\hat \Phi }}^H}{\mathbf{n}}_r\). Moreover, \({\mathbf{n}}\) has the same statistical property as \(\mathbf{n}_r\).

Afterwards, a MIMO equalizer should be introduced to mitigate the negative influence from ISI. We only update the coefficients at the known pilot symbols \(\mathbf{s}_p\). Moreover, we modified the reference information symbols of our MIMO equalizer to
  \begin{equation}\label{equ:p.4}
  {\mathbf{I}}_{eq} = {\mathbf{\hat Hs}_p},
  \end{equation}
which is different from the reference information symbols \({\mathbf{I}}_{eq} = {\mathbf{s}_p}\) in the conventional MIMO equalizers. By doing so, the ISI can be mitigated without decoding the symbols, and an arbitrary MIMO decoding algorithm, such as SIC, can be applied after the MIMO equalization.

\subsection{Successive Interference Cancellation}
\label{sec:sic}
To focus on the \(k^{th}\) transmit channel, we can rewrite \eqref{equ:p.3} as
  \begin{equation}\label{equ:sic.1}
  {\mathbf{y}} = {\mathbf{Hs}} + {\mathbf{n}} = \sum\limits_{i = 1}^{{N_t}} {{{\mathbf{h}}_i}{s_i}} + {\mathbf{n}} = {{\mathbf{h}}_k}{s_k} + \sum\limits_{\begin{subarray}{l} i = 1 \\ i \ne k \end{subarray} }^{{N_t}} {{{\mathbf{h}}_i}{s_i}} + {\mathbf{n}},
  \end{equation}
where \({\mathbf{h}}_i\) is the \(i^{th}\) column vector of the channel matrix \(\mathbf{H}\), and \(s_i\) is the \(i^{th}\) element of the transmitted signal vector \(\mathbf{s}\). For the \(k^{th}\) channel, the second term in the last equation is referred to as the inter-channel interference (ICI). We may notice that the ICI can be reduced if we can properly cancel out some of the interfering channels.

\begin{figure}[tb]
  \centering
  \includegraphics{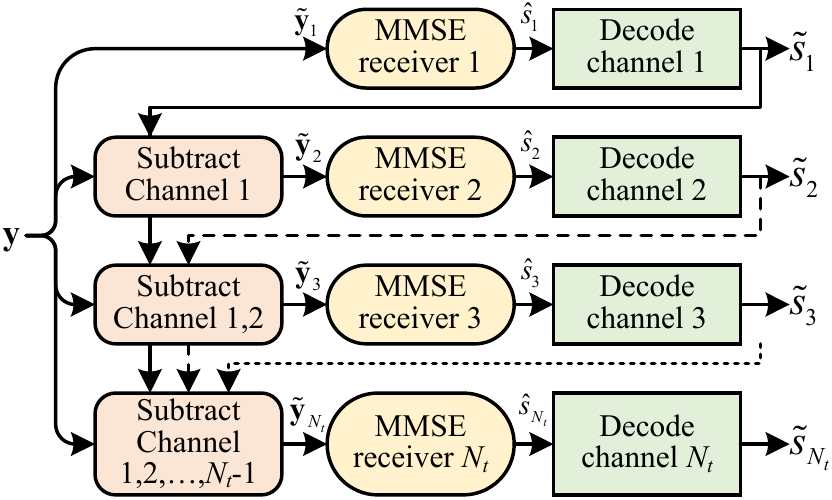}
  \caption{Schematic diagram for the SIC algorithm.}
  \label{fig:mmse_sic}
  \end{figure}

The schematic diagram for the SIC algorithm is shown in Fig.~\ref{fig:mmse_sic}.
The strategy of the SIC algorithm is to successively cancel out the interference of channels which have already been decoded \cite{tse2005fundamentals}. For the \(k^{th}\) channel, the subtraction process can be written as
  \begin{equation}\label{equ:sic.2}
  {{\mathbf{\tilde y}}_k} = {\mathbf{y}} - \sum\limits_{i = 1}^{k - 1} {{{{\mathbf{\hat h}}}_i}{{\tilde s}_i}},
  \end{equation}
where \({{{\mathbf{\hat h}}}_i}\) is the \(i^{th}\) column vector of the estimated channel matrix \(\mathbf{\hat H}\), and \({{\tilde s}_i}\) is the decoded symbol in the \(i^{th}\) channel. If the channel matrix is accurately estimated and the first \(k-1\) symbols are properly decoded, \eqref{equ:sic.2} can be rewritten as
  \begin{equation}\label{equ:sic.3}
  {{\mathbf{\tilde y}}_k} \approx {{\mathbf{h}}_k}{s_k} + \sum\limits_{i = k + 1}^{{N_t}} {{{\mathbf{h}}_i}{s_i}} + {\mathbf{n}},
  \end{equation}
where the approximation indicates that the nonlinear error propagation may occur in this process if estimation or decoding error occurs. However, the impact of this phenomenon can be minimized by using an optimal decoding order for each frame of the signal \cite{benesty2003fast}. Moreover, it will be shown in Section~\ref{sec:er} that we can still achieve a significantly better decoding performance by using the SIC algorithm.

After the subtraction process, the first \(k-1\) channels are cancelled out. Therefore, the \(k^{th}\) MMSE receiver should be modified to
  \begin{equation}\label{equ:sic.4}
  {\hat s_k} = {\left[ {{{\left( {{{{\mathbf{\hat H}}}_k}{\mathbf{\hat H}}_k^H + {N_0}{{\mathbf{I}}_{{N_r} \times {N_r}}}} \right)}^{ - 1}}{{{\mathbf{\hat h}}}_k}} \right]^H}{{\mathbf{y}}_k},
  \end{equation}
where \({{\mathbf{\hat H}}_k} = \left[ {{{{\mathbf{\hat h}}}_k},{{{\mathbf{\hat h}}}_{k + 1}}, \cdots ,{{{\mathbf{\hat h}}}_{{N_t}}}} \right]\), and \(\mathbf{I}_{{N_r} \times {N_r}}\) represents the \({N_r} \times {N_r}\) identity matrix.

Finally, decoding in the \(k^{th}\) channel can be calculated using the conventional decision rule as \cite{proakis2001digital}
  \begin{equation}\label{equ:sic.5}
  {\tilde s_k} = \mathop {\arg \min }\limits_{{{\tilde s}_k} \in \mathcal{S}} \left( {{{\left| {{{\tilde s}_k} - {{\hat s}_k}} \right|}}} \right),
  \end{equation}
where \(\mathcal{S}\) represents the set of all possible transmitted symbols, and \(\left| \cdot \right|\) is the modulus operator for complex numbers. 


\subsection{Vector Space Explanation}
\label{sec:vse}
\begin{figure}[htb]
  \centering
  \includegraphics{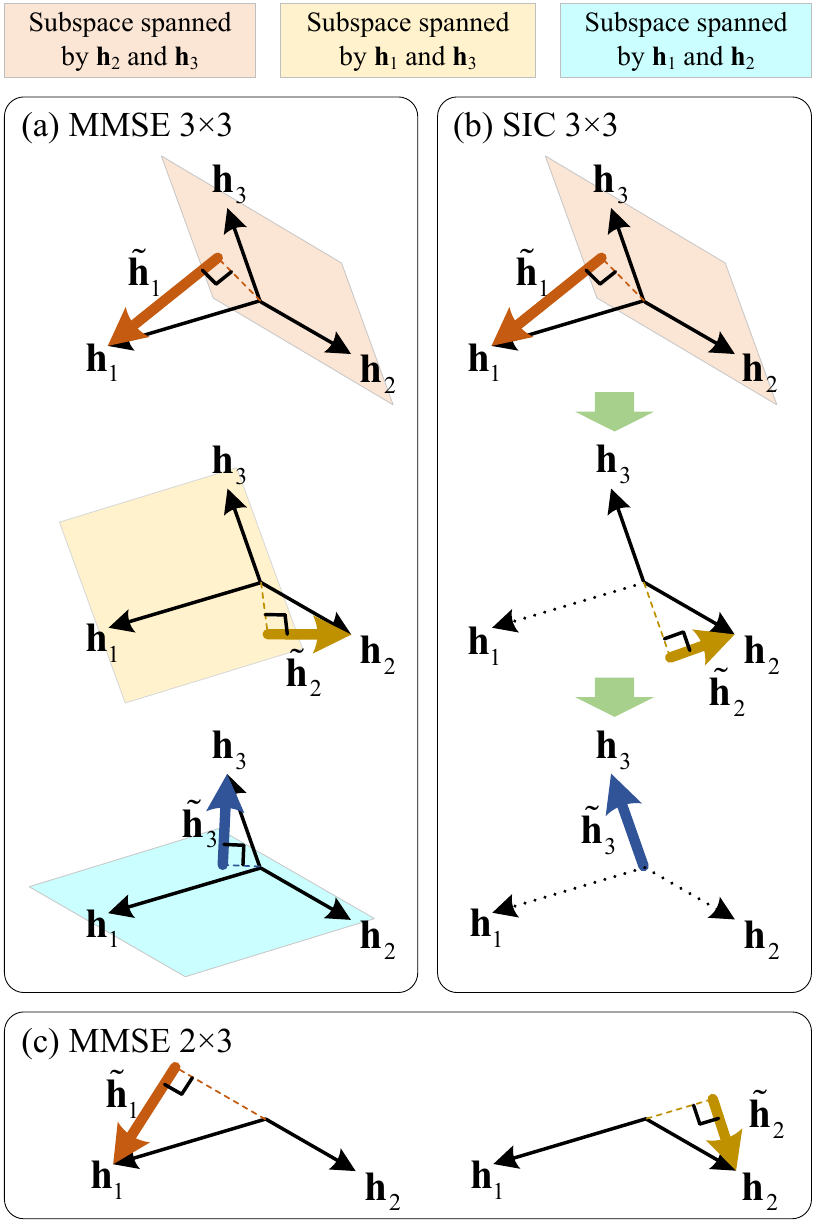}
  \caption{The vector space explanation for the enhanced atmospheric turbulence resiliency. \({{\mathbf{h}}_k}\) is the \(k^{th}\) column vector in the channel matrix \({\mathbf{H}}\), and \({{\mathbf{\tilde h}}_k}\) is the effective projection of \({{\mathbf{h}}_k}\). The vectors cancelled out by the SIC are represented by dotted vectors. (a) the \(3 \times 3\) MMSE MIMO decoder; (b) the \(3 \times 3\) SIC MIMO decoder; (c) the \(2 \times 3\) MMSE MIMO decoder with redundant receive channels.}
  \label{fig:sic_intuition}
  \end{figure}

In order to better understand why the proposed MIMO DSP algorithm can obtain enhanced atmospheric turbulence resiliency, Fig.~\ref{fig:sic_intuition} gives an vector space explanation for a system with 3 receive channels. In order to simplify the analysis, we only consider the special situation of negligible AWGN, where the MMSE algorithm degenerates to multiplying by a Moore–Penrose inverse matrix of \(\mathbf{H}\), which can also be regarded as a projection and a scaling of the column vectors in \(\mathbf{H}\). The AWGN in a practical system should be considered as another dimension in the random vector space and analogous analysis can then apply to the generalized situation with AWGN.

Fig.~\ref{fig:sic_intuition}(a) illustrates a \(3 \times 3\) linear MMSE MIMO decoder in vector space. Considering the fact that the scaling effect is just for normalization, we are only interested in the projection effect. For arbitrary \({{\mathbf{h}}_k}\), the effective projection \({{\mathbf{\tilde h}}_k}\) should be in the 1 dimensional orthogonal complement of the subspace spanned by \(\left\{ {{{\mathbf{h}}_i}\left| {1 \leqslant i \leqslant {N_t},i \ne k} \right.} \right\}\) to eliminate the negative influence of ICI.

Fig.~\ref{fig:sic_intuition}(b) illustrates a \(3 \times 3\) SIC MIMO decoder. In the first channel, no interference cancellation is performed and the same performance as the MMSE decoder should be expected. However, when decoding the second channel, the first channel can be cancelled out, and \({{\mathbf{\tilde h}}_2}\) is a projection in the orthogonal complement of \({{\mathbf{h}}_3}\), which is a 2 dimensional vector subspace. Therefore, the degree of freedom is increased to 2 for the second channel. Similarly, the last channel needs no projection and all the 3 degrees of freedom can be exploited. As a result, the effective projections in the SIC system will be longer than or equal to the effective projections in the MMSE system, and a larger effective signal-to-interference-plus-noise ratio (SINR) can be achieved in the SIC system.

Fig.~\ref{fig:sic_intuition}(c) illustrates a \(2 \times 3\) MMSE MIMO decoder which has redundant receive channels. The third channel is directly cancelled out because we are not transmitting any symbols in that channel. As a result, arbitrary \({{\mathbf{h}}_k}\) can always be projected onto a 2 dimensional orthogonal complement subspace and a better turbulence resiliency can be obtained. 

It is also worth noting that in an ideal system with mutually orthogonal column vectors in the channel matrix, the equation \({{\mathbf{\tilde h}}_k} = {{\mathbf{h}}_k}\) holds. Therefore, the system performance can not be improved by either applying the SIC algorithm or adding redundant receive channels. Unfortunately, although the orthogonality can be well conserved in certain kinds of channels (e.g. short FMF link or back-to-back FSO link without turbulence), this condition doesn't hold in a turbulent channel.

\section{Experimental Results}
\label{sec:er}

Table~\ref{tab:er.1} lists the experimental parameters for our MDM FSO communication system. 
\begin{table}[htb]
  \caption{Experimental Parameters}
  \label{tab:er.1}
    \begin{center} 
      \renewcommand{\arraystretch}{1.3}
      \begin{tabular}{|c|c|}
        \hline
        Parameter & Value \\
        \hline
        Wavelength & 1550.12\,nm \\
        \hline
        Laser linewidth & 100\,kHz \\
        \hline
        Symbol format & DP-QPSK \\
        \hline
        Baud rate & 34.46\,GBaud \\
        \hline
        Roll-off factor & 0.1 \\
        \hline
        Receiver sampling rate & 50\,GSa/s \\
        \hline
        \multicolumn{1}{|m{4.9cm}|}{\centering Hard-decision forward error correction \\ (HD-FEC) limit with 6.25\% overhead \cite{zhang2014staircase}} & \multicolumn{1}{m{1.5cm}|}{\centering \( 4.7 \times 10^{-3} \)}\\
        \hline
        Diameter of the receiver lens (D) & 8.4\,mm \\
        \hline
        Fried's parameter (\(r_0\)) for weaker turbulence & 3.0\,mm \\
        \hline
        Fried's parameter (\(r_0\)) for stronger turbulence & 0.8\,mm \\
        \hline
        Inner scale of turbulence (\(l_0\)) & 0.1\,mm\\
        \hline
        Outer scale of turbulence (\(L_0\)) & 10\,m\\
        \hline
      \end{tabular}
    \end{center}
  \end{table}

\begin{figure}[tb]
  \centering
  \includegraphics{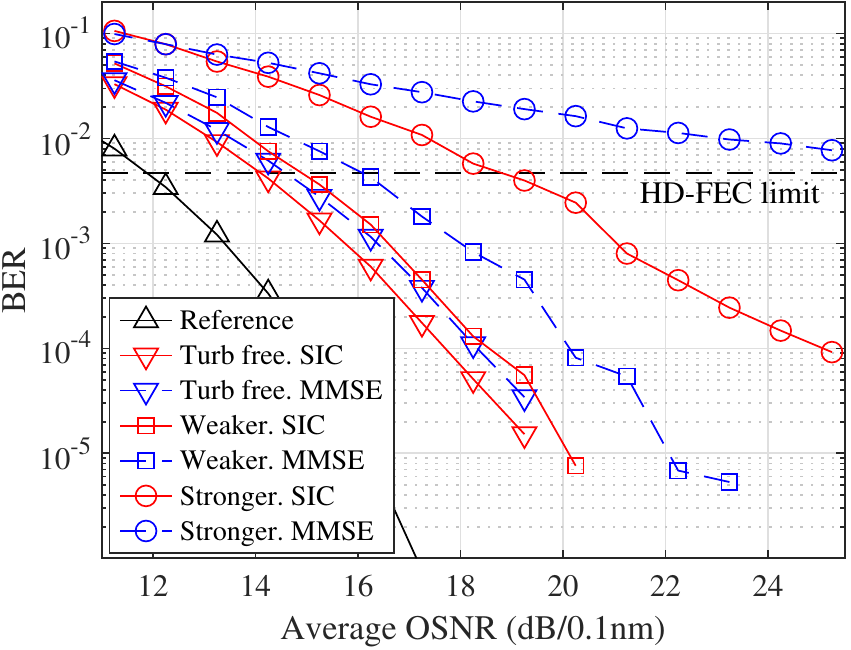}
  \caption{Average BER curves of the \(10 \times 12\) MIMO system under different turbulence intensities. Reference: Theoretical reference curve in an ideal system; Turb free: turbulence free system with a blank pattern. Weaker: typical weaker turbulence pattern when \(r_0 = 3.0~\text{mm}\); Stronger: typical stronger turbulence pattern when \(r_0 = 0.8~\text{mm}\).}
  \label{fig:PlotBerOsnr_5x6}
  \end{figure}

First, we examine the BER performance against average OSNR of the \(10 \times 12\) MIMO system under different turbulence conditions, which is shown in Fig.~\ref{fig:PlotBerOsnr_5x6}. 
The theoretical reference curve assumes that: 
(1) The channel matrix is a $N_r \times N_t$ submatrix of a $N_r \times N_r$ unitary matrix;
(2) No other imperfection except the AWGN exists.
For the turbulence free situation, a blank pattern with factory pre-calibration is used in the system, here observe a \textasciitilde 2.2\,dB implementation penalty for the SIC system at the hard-decision forward error correction (HD-FEC) limit, which we believe is dominated by the inter-mode crosstalk and residual mode-dependent losses in the MSPL pair.
As expected, these imperfections have a slightly higher impact on the MMSE system (\textasciitilde 2.7\,dB implementation penalty).
In the weaker turbulence case, the SIC system has a \textasciitilde 3\,dB penalty but the penalty for the MMSE system is \textasciitilde 4.2\,dB, which indicates a moderate reduction in channel matrix orthogonality.
For the stronger turbulence, a larger penalty of \textasciitilde 6.9\,dB 
is observed in the SIC system, 
but the MMSE system does not even achieve the HD-FEC limit, and the loss of channel matrix orthogonality is severe.
The result in the stronger turbulence case also indicates that the proposed DSP algorithm has only mitigated rather than eliminated the negative influence of atmospheric turbulence.

\begin{figure}[tb]
  \centering
  \includegraphics{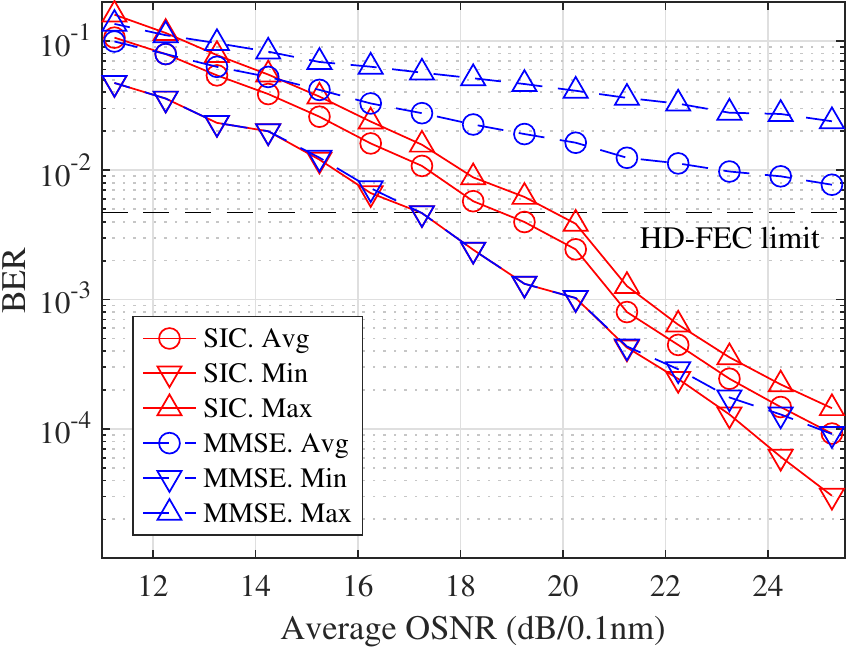}
  \caption{BER curves of the \(10 \times 12\) SIC (red lines) and MMSE (blue dotted lines) MIMO systems under the stronger turbulence when \(r_0 = 0.8\,\text{mm}\). Avg (circles): average BER; Min (lower triangles): the best channel with minimum BER at each average OSNR; Max (upper triangles): the worst channel with maximum BER at each average OSNR.}
  \label{fig:PlotBerOsnrDetail_5x6}
  \end{figure}

In order to better understand the enhanced turbulence resiliency by the proposed SIC system, Fig.~\ref{fig:PlotBerOsnrDetail_5x6} depicts the BER of the best channel (the channel with minimum BER at each average OSNR), the worst channel (the channel with maximum BER at each average OSNR), and the average BER of all channels (also shown in Fig.~\ref{fig:PlotBerOsnr_5x6}) in the stronger turbulence.
The best channels in both the SIC system and the MMSE system have similar BER performance, this is because we don't execute any interference cancellation for the first decoded channel in the SIC system. At the high OSNR region, the best channel in the SIC system performs slightly better than the MMSE system. This is because the second decoded channel may obtain a better BER performance by cancelling out the first decoded channel.
On the other hand, the worst channel in the SIC system has a much better performance than it in the MMSE system. This is because a higher effective SINR can be achieved in the SIC system by cancelling out the previously decoded channels. 
Considering the fact that the worst channel is the dominant factor in the average BER performance, the SIC system can obtain a significantly lower average BER in turbulent channels.

\begin{figure*}[htb]
  \centering
  \includegraphics[width=7.1in]{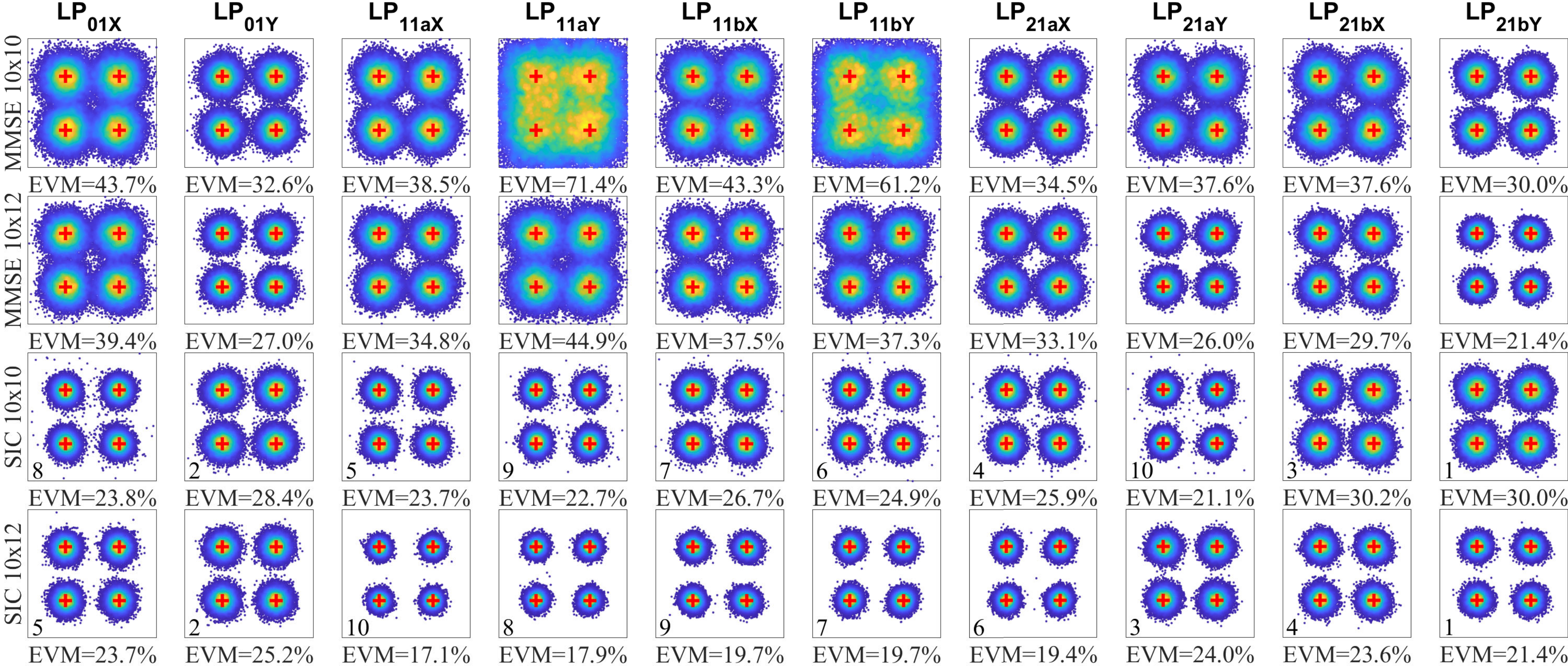}
  \caption{Constellation diagrams of different transmitted channels for the \(10 \times 10\) MMSE (the first row), \(10 \times 12\) MMSE (the second row), \(10 \times 10\) SIC (the third row), and \(10 \times 12\) SIC (the last row) MIMO configurations without noise loading. The SIC order is given at the bottom left of each constellation. EVM: error vector magnitude. Red crosses: reference QPSK constellation points.}
  \label{fig:constellation}
  \end{figure*}

The constellation diagrams for the \(10 \times 10\) and \(10 \times 12\) MIMO systems under the stronger turbulence case are depicted in Fig.~\ref{fig:constellation} without noise loading.
The SIC order is given at the bottom left of each constellation.
Because the channel matrices are different between the \(10 \times 10\) and \(10 \times 12\) MIMO systems, the SIC order may also be different.
As shown in Fig.~\ref{fig:constellation}, the \(10 \times 10\) and \(10 \times 12\) SIC systems have a significantly lower error vector magnitude (EVM) than the \(10 \times 10\) and \(10 \times 12\) MMSE systems, respectively. This is because certain ICI is cancelled out by applying the SIC algorithm. However, no ICI is cancelled out in the first decoded channel (\({\text{LP}_\text{21bY}}\) in both cases). In this channel, identical EVM performance is observed in both SIC and MMSE decoders.
On the other hand, a comparison between the \(10 \times 10\) and \(10 \times 12\) MMSE MIMO systems shows that 1 more redundant receive mode (2 more redundant receive channels) can significantly reduce the EVM of the constellation in all the 10 channels.
Compared with the \(10 \times 10\) SIC MIMO system, We can also obtain a better average EVM performance in the \(10 \times 12\) SIC MIMO system because of the information obtained from the 2 redundant receive channels (\({\text{EVM = 21}}{\text{.23\% }}\) for the \(10 \times 12\) SIC MIMO system and \({\text{EVM = 25}}{\text{.78\% }}\) for the \(10 \times 10\) SIC MIMO system, respectively),
although a higher EVM in specific channels (e.g. \({\text{LP}_\text{21aY}}\)) may be observed due to the different SIC order.
Moreover, less nonlinear propagation (apparently discrete points in the diagram, see for example \({\text{LP}_\text{01X}}\)) is also observed in the \(10 \times 12\) SIC MIMO system. This is also because of the information obtained from the 2 redundant receive channels and less symbol error occurs in the decoded channels.

\begin{figure}[tb]
  \centering
  \includegraphics{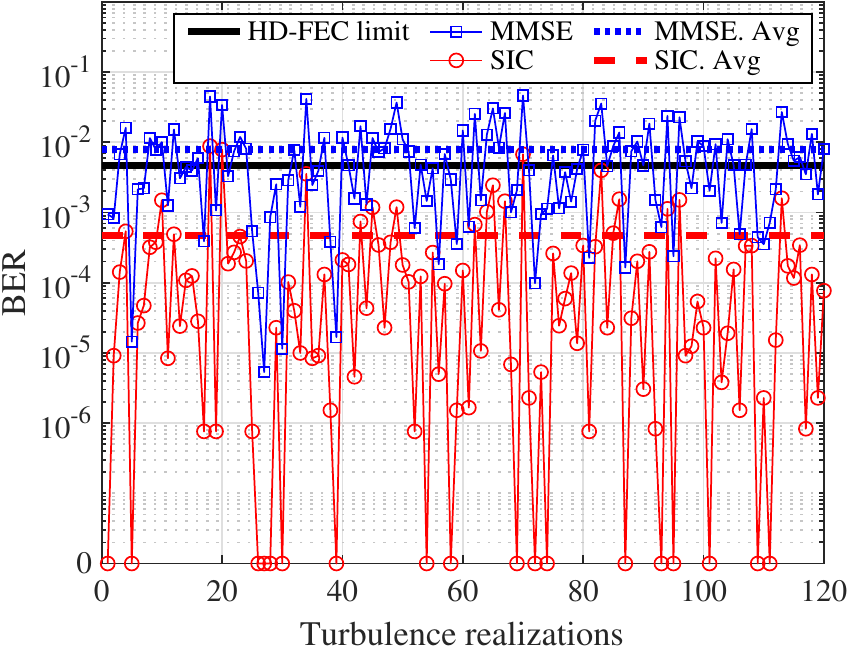}
  \caption{The BER performance of the \(10 \times 12\) MIMO system under 120 stronger turbulence realizations.}
  \label{fig:PlotBerScan_5x6}
  \end{figure}
\begin{figure}[tb]
  \centering
  \includegraphics{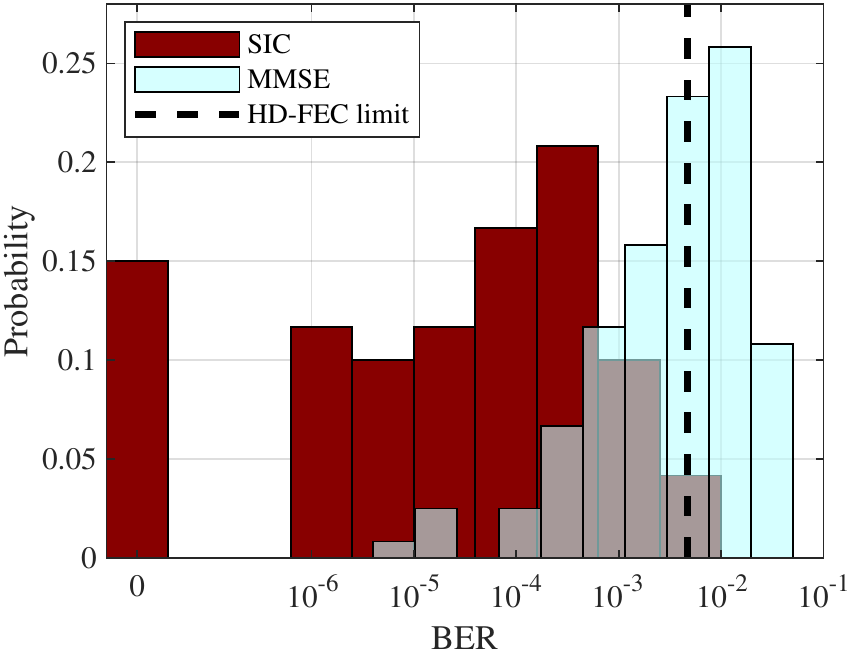}
  \caption{The probability distribution of the BER performance in the \(10 \times 12\) system under 120 stronger turbulence realizations.}
  \label{fig:PlotBerScanPdf_5x6}
  \end{figure}

We tested 120 independent turbulence patterns under the stronger turbulence case without noise loading. As shown in Fig.~\ref{fig:PlotBerScan_5x6}, the BER of the SIC system is consistently lower than the MMSE system. 
As a result, the average BER among different turbulence realizations is decreased by an order of magnitude, from \(8.02 \times 10^{-3}\) in the MMSE system to \(4.76 \times 10^{-4}\) in the SIC system (well below the HD-FEC limit).
The probability distribution of the BER in Fig.~\ref{fig:PlotBerScan_5x6} is depicted in Fig.~\ref{fig:PlotBerScanPdf_5x6} for better clarity.
Because of the limited length of our recorded data, the BER should be taken as being below \(10^{-6}\) for the 18 turbulence realizations where no errors we recorded.
If we consider an outage when the BER is larger than the HD-FEC limit, we have an outage probability of 48.3\% 
in the MMSE system and only 2.5\% 
in the SIC system, suggesting a more robust transmission using SIC.

\begin{figure}[tb]
  \centering
  \includegraphics{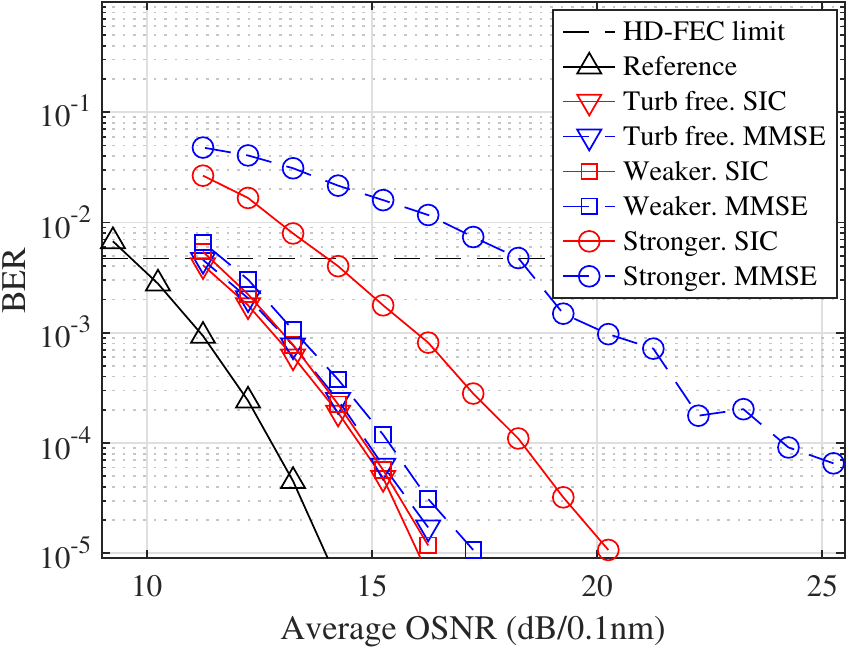}
  \caption{Average BER curves of the \(6 \times 12\) MIMO system under different turbulence intensities. Reference: Theoretical reference curve in an ideal system; Turb free: turbulence free system with a blank pattern. Weaker: typical weaker turbulence pattern when \(r_0 = 3.0\,\text{mm}\); Stronger: typical stronger turbulence pattern when \(r_0 = 0.8\,\text{mm}\).}
  \label{fig:PlotBerOsnr_3x6}
  \end{figure}

To further illustrate the enhanced turbulence resiliency obtained from the redundant receive channels, we increased the ratio of received to transmitted channels by disabling the two highest order mode inputs of the transmitter. Fig.~\ref{fig:PlotBerOsnr_3x6} depicts the BER performance against average OSNR of a \(6 \times 12\) MIMO system under the same turbulence conditions as Fig.~\ref{fig:PlotBerOsnr_5x6}.
For the turbulence free situation, a blank SLM pattern with factory pre-calibration is used in the system.
In this turbulence free scenario, we observe a \textasciitilde 1.4\,dB implementation penalty between the reference and the SIC system at the HD-FEC limit, while the MMSE system has a slightly higher implementation penalty of \textasciitilde 1.5\,dB.
In the weaker turbulence case, the performance of the SIC system has a \textasciitilde 1.7\,dB implementation penalty, compared to \textasciitilde 2.0\,dB for the MMSE system.
In the stronger turbulence case, a larger implementation penalty of \textasciitilde 4.0\,dB is observed in the SIC system, while the implementation penalty of the MMSE system is \textasciitilde 8.2\,dB.
When compared with the corresponding results in Fig.~\ref{fig:PlotBerOsnr_5x6}, we can conclude that a significantly better turbulence and device imperfection resiliency can be obtained under different turbulence conditions by increasing the number of redundant receive channels in the MDM FSO system.
  
\begin{figure}[tb]
  \centering
  \includegraphics{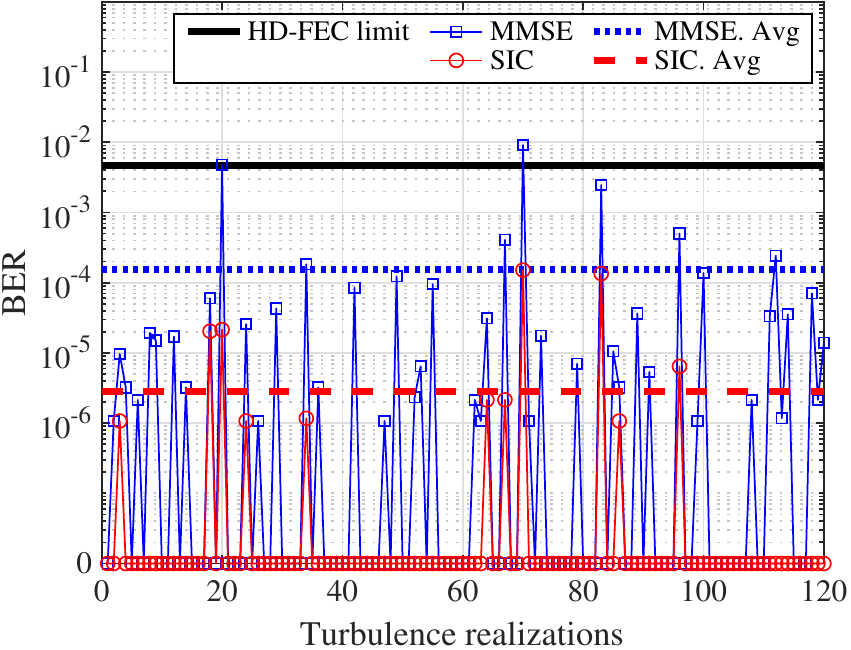}
  \caption{The BER performance of the \(6 \times 12\) MIMO system under 120 stronger turbulence realizations.}
  \label{fig:PlotBerScan_3x6}
  \end{figure}

Fig.~\ref{fig:PlotBerScan_3x6} depicts the BER performance of a \(6 \times 12\) system using 120 independent patterns, which are the same as the patterns used in Fig.~\ref{fig:PlotBerScan_5x6}, under the stronger turbulence case without noise loading.
Similar to Fig.~\ref{fig:PlotBerScan_5x6}, the BER of the SIC system is consistently lower than the MMSE system. As a result, the average BER among different turbulence realizations is decreased from \(1.56 \times 10^{-4}\) to \(2.86 \times 10^{-6}\). If we consider an outage when the BER is larger than the HD-FEC limit, the outage probability of the MMSE and the SIC system will be decreased to 1.67\%
and 0\%,
respectively. When compared with the corresponding results in Fig.~\ref{fig:PlotBerScan_5x6}, we can conclude that a significantly better turbulence resiliency can be obtained in different turbulence realizations by increasing the number of redundant receive channels.


  
\begin{figure}[tb]
  \centering
  \includegraphics{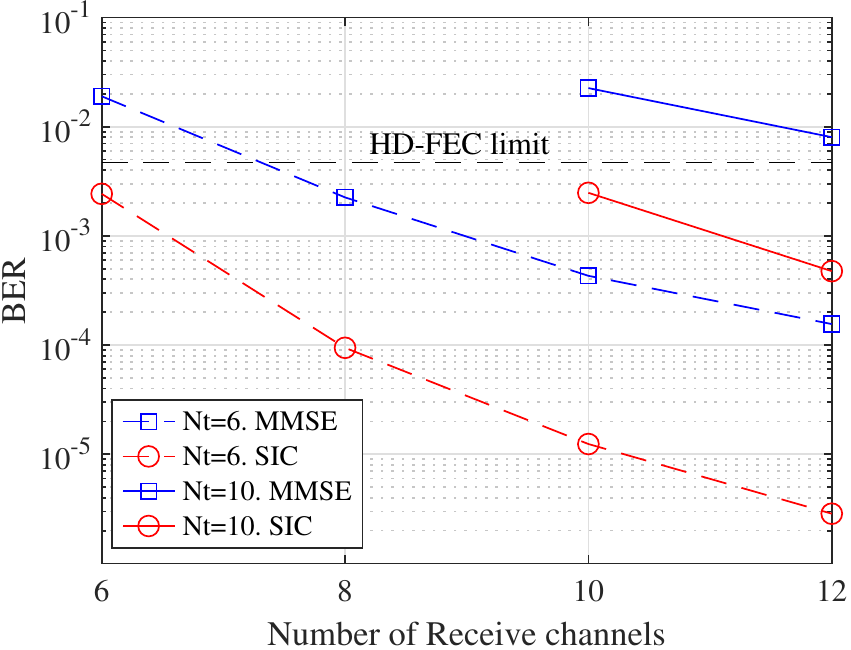}
  \caption{The average BER performance of different MIMO systems under 120 stronger turbulence realizations.}
  \label{fig:PlotBerTxRx}
  \end{figure}

In order to comprehensively compare different transmit and receive channel numbers as well as different MIMO decoding algorithms, Fig.~\ref{fig:PlotBerTxRx} depicts the average BER performance of different MIMO systems using 120 independent patterns under the strong turbulence case without noise loading.
As shown in Fig.~\ref{fig:PlotBerTxRx}, when the number of transmit channels (\(N_t\)) is fixed, the average BER will decrease as the number of receive channels (\(N_r\)) increases. On the other hand, when \(N_r\) is fixed, the average BER will increase as \(N_t\) increases. Moreover, in the same system setup when \(N_t\) and \(N_r\) are both fixed, the SIC algorithm always performs better than the MMSE algorithm.

\section{Conclusion}
\label{sec:c}
In this paper, we experimentally demonstrate the enhanced turbulence resiliency in a MDM FSO communication link with 5 transmitted spatial modes, each carrying a 34.46\,GBaud DP-QPSK signal.
By employing the SIC MIMO algorithm and redundant receive channels, the average BER of the \(10 \times 12\) MIMO system goes down to \(4.76 \times 10^{-4}\), which is well below the HD-FEC limit, among 120 independent turbulence realizations where $D=8.4$\,mm and \(r_0=0.8\,\text{mm}\) (\(D/r_0=10.5\)).
Moreover, the outage probability is significantly decreased when using the SIC algorithm, which indicates the possibility of a much more robust transmission in the MDM FSO system through turbulent channels.

We also compare the system performance among systems with different transmit and receive channel numbers. The results indicate that we can obtain an even better BER performance by
increasing the number of redundant receive channels. These results may also provide useful guidelines when deploying the MDM FSO system in an even stronger turbulence environment.

As a result, the system demonstrates a record-high independent channel number of 10 and line rate of 689.23\,Gbit/s. Considering the 8.4\% training sequence, 10\% pilot rate, 6.25\% HD-FEC cost, and 0.1 rolloff factor, a record-high net spectral efficiency of 13.9\,bit/s/Hz was also achieved in emulated turbulent MDM FSO links.


%




\section*{Acknowledgment}

We thank Dr. Chao Gao for fruitful discussions. We also thank Ciena and Dr. Charles Laperle for kindly providing the WaveLogic 3 transponder used in our experiments.




\ifCLASSOPTIONcaptionsoff
  \newpage
\fi



\bibliographystyle{IEEEtran}
\bibliography{paperbib}

\end{document}